\documentclass[twocolumn,aps,floats,showpacs,superscriptaddress,floatfix]{revtex4}
\usepackage[final]{epsfig}
\usepackage{amsmath, amsthm, amssymb}

\newcommand{\Avg}[1]{\left\langle{#1}\right\rangle}
\newcommand{\be}{\begin{equation}}
\newcommand{\ee}{\end{equation}}
\newcommand{\beas}{\begin{eqnarray*}}
\newcommand{\eeas}{\end{eqnarray*}}
\newcommand{\bea}{\begin{eqnarray}}
\newcommand{\eea}{\end{eqnarray}}

\newcommand{\wt}{\widetilde}

\begin{document} 
\title{Mixing properties of growing networks and the Simpson's paradox}
\author{Andrea Capocci}
\affiliation{Centro Studi e Ricerche E. Fermi, Compendio
  Viminale, Roma, Italy}
\author{Francesca Colaiori} 
\affiliation{Dipartimento Fisica, 
Universit\`a di Roma ``La Sapienza" and SMC, INFM UdR Roma1 
Piazzale Aldo Moro 2, 00185, Roma, Italy}
\date{\today}
\begin{abstract}
We analyze the mixing properties of growing networks and find that, in
some cases, the assortativity patterns are reversed once links'
direction is considered: the disassortative behavior observed in such
networks is a spurious effect, and a careful analysis reveals genuine
positive correlations. We prove our claim by analytical calculations
and numerical simulations for two classes of models based on
preferential attachment and fitness. Such counterintuitive phenomenon
is a manifestation of the well known Simpson's paradox. Results
concerning mixing patterns may have important consequences, since they
reflect on structural properties as resilience, epidemic spreading and
synchronization. Our findings suggest that a more detailed analysis
of real directed networks, such as the World Wide Web, is needed.
\end{abstract}
\maketitle 

Complex networks arise in a wide range of interacting structures,
including social, technological and biological systems
\cite{NetReviews}. Although all these networks share some generic
statistical features, such as the small world property and the
scale--invariance of the degree distribution, they also display
differences and peculiarities when their structure is examined in
detail.

A distinctive characteristic of a network is whether its nodes tend to
connect to similar or unlike peers, the so--called mixing property
\cite{newman02}. Similarity of nodes is established by comparing some
node--dependent scalar quantity measuring a given quality. Borrowing
terms from sociology, networks where properties of neighboring nodes
are positively correlated are called {\it assortative}, while those
showing negative correlations are called {\it disassortative}. Thus,
assortative and disassortative mixing patterns indicate a generic
tendency to connect respectively to similar or dissimilar pears. A
scalar quantity naturally associated to each node in a network is its
degree, measuring the number of neighboring nodes. The mixing by
degree (MbD) is often measured by looking at how the average degree
$K_{nn}$ of the nearest neighbors of a node depends on the degree $K$
of the node itself, and is a signature of correlations between other
networks quantities \cite{fronczak05}. The mixing is assortative
when $K_{nn}$ grows with $K$ and disassortative 
when it decreases \cite{pastor01}.
The relevance of MbD lies in that, beyond discriminating among
different network morphologies \cite{palla05}, it reflects important
structural properties. Assortative networks are found to be more
resilient against the removal of vertexes than disassortative ones
\cite{vazquez03}. This implies, for example, that, when trying to
block infection or opinion spreading within a social network
\cite{hayashi04,blanchard05}, or to protect a computer network against
cyber--attacks \cite{hayashi05}, different strategies are needed
depending on the MbD properties of the underlying network. Moreover,
it has recently been observed that the sign of degree correlations
affect other properties of complex networks such as synchronization
\cite{bernardo05}.

Recent studies show that social networks exhibit assortative MbD,
whereas technological and biological ones display disassortative MbD
\cite{newman03b}. The Word Wide Web (WWW), a paradigmatic example of
world--wide collaborative effort among millions of users and
publishers, represents an anomaly: one would expect it to show
assortative mixing, similarly to other social and collaborative
networks, while it shows evidences of anticorrelations
\cite{newman02}, and disassortative MbD \cite{capocci03}, which would
rather put it in the realm of technological networks.

We aim to show that, in networks where a direction is naturally
associated to the links, like in growing networks, it is crucial to
distinguish between nearest neighbors along incoming and outgoing
links. In the WWW case, for example, links with different direction
have different roles and meanings: the outgoing links are drawn by
individual web--masters, while they have no control on incoming
links. In the language of Kleinberg \cite{kleinberg00} a page gains
authority from incoming links, while it increases a peer's authority
by pointing to it. Nevertheless, the WWW has been often analyzed and
modeled as an undirected network for what concerns its mixing
properties \cite{capocci03,newman03c}.

Our main result is that, in most cases, assortativity patterns are
reversed when the direction of links in a network is taken into
account: positive correlations among the degree of a node and the
average degree of both upstream and downstream neighbors, considered
separately, can disappear or even reverse when the different nature of
neighboring sites is ignored and their degree are averaged together.
Though this result may appear counterintuitive, the fact that pooling
together data of different nature can generate spurious correlations
is well known in the statistical literature, and often encountered in
social sciences, medical statistics and finance, where, although it
contains no logical contradiction, it is known as {\em Simpson's
paradox} \cite{simpson51}.

We show our result on two classes of complex growing networks: the
linear preferential attachment (LPA) model \cite{dorogotsev00}, and
the Bianconi--Barabasi (BB) fitness model \cite{bianconi01}. Both
include as a special case the Barabasi--Albert (BA) model
\cite{barabasi99}. In such growing network models, links have a
natural direction -- from newly added nodes to existing ones. Thus,
in the following we distinguish between upstream and downstream
neighbors, respectively along incoming and outgoing links.

\begin{figure}
\centerline{\psfig{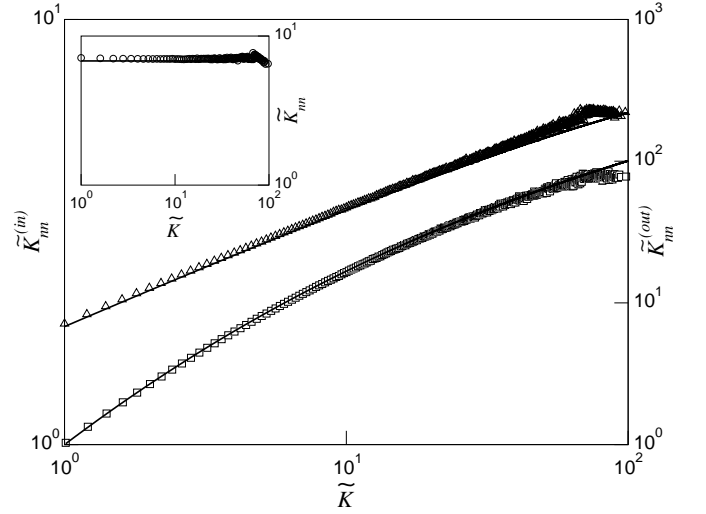}} 
\caption{Histogram of $\wt{K}_{nn}^{(in)}$ (squares),
$\wt{K}_{nn}^{(out)}$(triangles), as functions of $\wt{K}$ from
simulations of the the BA model, with $m=100$, $t=10^4$, and averaged
over $10^4$ realizations.The behavior of $\wt{K}_{nn}$ is shown in the
inset. The solid lines are the analytic calculations.}
\label{BA}
\end{figure}

To clarify our argument, we consider in detail the BA model (where
calculations are simpler) before moving to the LPA and BB models.  In
the BA model, at each time step a node is added and attached to the
network by $m$ undirected links with preferential attachment. A node
$i$ (introduced at time $i$) points to existing nodes $j$ with
probability $p_j(i)$ proportional to their degree $K_j(i)$ at time $i$
\cite{barabasi99}. Since $m$ sets a natural scale for the system, we
will express all quantities in units of $m$, and denote them with the
superscript $\sim$.  On average, the degree of node $i$ grows in time
as $\wt{K}_i(t)\simeq \sqrt{t/i}$ for $1\ll i\ll t$.  The average
degrees of neighbors of $i$, in $m$ units, read $
\wt{K}_{nn,i}^{(in)}(t) =\sum_{j=i+1}^t
\wt{K}_j(t)\wt{p}_i(j)/(\wt{K}_i(t)-1)$ and
$\wt{K}_{nn,i}^{(out)}(t)=\sum_{j=1}^{i-1} \wt{K}_j(t)\wt{p}_j(i) $,
where $\wt{K}_{nn,i}^{(in)}(t)$ and $\wt{K}_{nn,i}^{(out)}(t)$ refer
to the degree of upstream and downstream neighbors respectively.  By
approximating the sum by an integral and the degree by its average,
one gets $\wt{K}_{nn,i}^{(in)}(t)\simeq\log{\sqrt{t/i}}
/(1-\sqrt{i/t})$ and $\wt{K}_{nn,i}^{(out)}(t)\simeq
\sqrt{t/i}\log(A\sqrt{i})$, where $A$ is a constant of order one whose
exact value depends on the initial condition \cite{nota1}.  At a given
time $t$, we can express the above quantities in terms of $\wt{K}$
and drop the $i$ dependence to get $
\wt{K}_{nn}^{(in)}\simeq\wt{K}\log{\wt{K}} /(\wt{K}-1)$ and
$\wt{K}_{nn}^{(out)}\simeq \wt{K}\log (\overline{K}/\wt{K})$, where
$\overline{K}=A\sqrt{t}$ is of order of (and greater than) the maximum
$\wt{K}$ observable at time $t$, which is
$\wt{K}_{max}\simeq\sqrt{t}$.  Thus $\wt{K}_{nn}^{(in)}$ is a
monotonically (slowly) increasing function of $\wt{K}$, independent on
$t$, and $\wt{K}_{nn}^{(out)}$ contains a $t$ dependence through
$\overline{K}$ and for any $t$ is an increasing function of
$\wt{K}$.  We conclude that the degree of a node is positively
correlated both with the average degree of upstream and downstream
neighbors.
However, computing the average degree of the neighbors altogether,
correlations are lost and one gets $ \wt{K}_{nn}(t)\simeq \log
(A\sqrt{t})$, independent on $\wt{K}$ \cite{boguna03}. These results
are confirmed by numerical simulation of the BA model and shown in
Fig.\ref{BA}, where histograms of $\wt{K}_{nn}^{(in)}$,
$\wt{K}_{nn}^{(out)}$, and $\wt{K}_{nn}$ are plotted as functions of
$\wt{K}$ for $t=10^4$ and $m=100$, averaged over $10^4$ realizations.

Let us now focus on the LPA model \cite{dorogotsev00}, a
generalization of the BA model: according to the same dynamics, at the
$i$--th time step $m$ directed links are drawn from $i$ to $j$ with
probability $p_j(i) \propto k_j(i)+\alpha$, where $k_j(i)$ is the
in--degree of site $j$ at time $i$.  For $\alpha = m$, the BA model is
recovered. When dealing with the LPA model, it is convenient to
measure quantities in units $\alpha$. In the continuum time limit, the
time dependence of the in--degree is $\wt{k}_i(t)=(t/i)^{\beta}-1$ with
$\beta=(1+\alpha/m)^{-1}$. The degrees are power--law distributed with
exponent $\gamma = 2+\alpha/m$ \cite{dorogotsev00}.
The calculation of the average in--degree of upstream and downstream
neighbors can be performed in analogy to the BA model. The average
degree of upstream neighbors reads
$\wt{k}_{nn}^{(in)}\simeq(\wt{k}+1)\log(\wt{k}+1)/\wt{k}-1$, as in the
BA model since it is independent from the ratio $\alpha/m$, and is
monotonically increasing. The average degree of downstream neighbors
is given by 
\be \wt{k}_{nn}^{(out)}\simeq
(1-\beta)(\wt{k}+1) \frac{\left(
\frac{\bar{k}+1}{\wt{k}+1}\right)^{\frac{2\beta-1}{\beta}}-1}
{2\beta-1}-1 \,,
\label{knnoutLPA}
\ee 
which is also an increasing function of the in--degree $\wt{k}$.
$\wt{k}_{nn}^{(out)}$ contains an explicit dependence on $\alpha /m$
through $\beta$ and on $t$ through $\bar{k}$ ($\bar{k}\gtrsim
\wt{k}_{max}\simeq t^{\beta}$).  Note that now $\wt{k}_{nn}^{(in)}$
and $\wt{k}_{nn}^{(out)}$ count incoming links only.  Instead, when
ignoring the direction of links by averaging the degree over all
nodes' neighbors, one gets 
\be 
\wt{k}_{nn} \simeq
\frac{\wt{k}+1}{\wt{k}+\frac{\beta}{1-\beta}} \left( \log(\wt{k}+1)+
\frac{
\left(\frac{\bar{k}+1}{\wt{k}+1}\right)^{\frac{2\beta-1}{\beta}}-1}
{2\beta-1} \right)-1 \,.
\label{knnLPA}
\ee Two different regimes appear, for $\alpha/m<1$ ($\beta>1/2$) and
$\alpha/m>1$ ($\beta<1/2$), separated by $\alpha=m$ where the LPA
model coincides with the BA model.  The average in--degree of nearest
neighbors increases as a function of $\wt{k}$ for $\beta<1/2$, while
it decreases for $\beta > 1/2$ \cite{barrat05}. The two regimes
correspond to qualitatively different behaviors of the degree
distribution: for $\alpha/m>1$ the distribution has finite variance in
the thermodynamic limit($\gamma>3$), while $\alpha/m<1$ corresponds to
$2<\gamma<3$, with diverging variance in the same limit.  Summarizing,
for the LPA model the degree of a node is positively correlated with
the average degree of both upstream and downstream nearest neighbors.
However, the average degree over all nearest neighbors increases or
decreases for different values of the parameter $\beta$. A behavior
similar to the $\beta > 1/2$ case was already observed in simulations
of a weighted directed model for the WWW \cite{barrat04}.
In Fig 3 and
4 we show the results of our calculation, compared with simulation of
the LPA model for $m=100$, $t=10^4$, and $\alpha=5$ for the
$\beta>1/2$ regime (Fig. 3), and $m=100$, $t=10^4$, and $\alpha=500$
for the $\beta<1/2$ regime (Fig. 4).

\begin{figure}
\centerline{\psfig{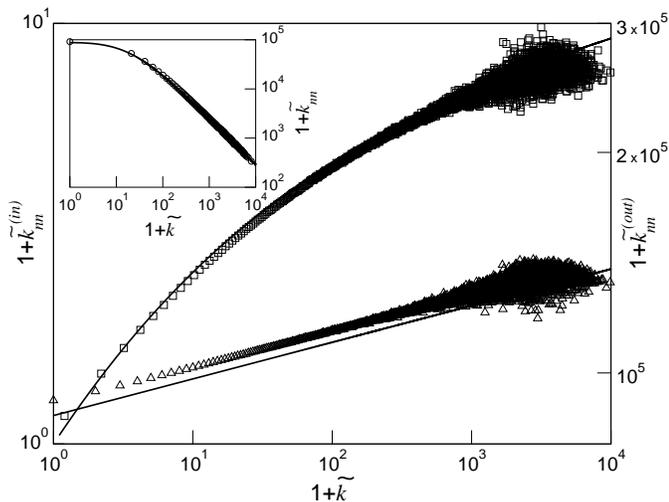} }
\caption{Histogram of $\wt{k}_{nn}^{(in)}$ (squares),
$\wt{k}_{nn}^{(out)}$ (triangles), as functions of $\wt{k}$ for
$t=10^4$, from simulations of the LPA model with $m=100$ and
$\alpha=5$ ($\beta>1/2$), and averaged over $10^4$ realizations. The
behavior of $\wt{k}_{nn}$ is shown in the inset. The solid lines are
the analytic calculations.}
\label{BA_Knnk}
\end{figure}
\begin{figure}
\centerline{\psfig{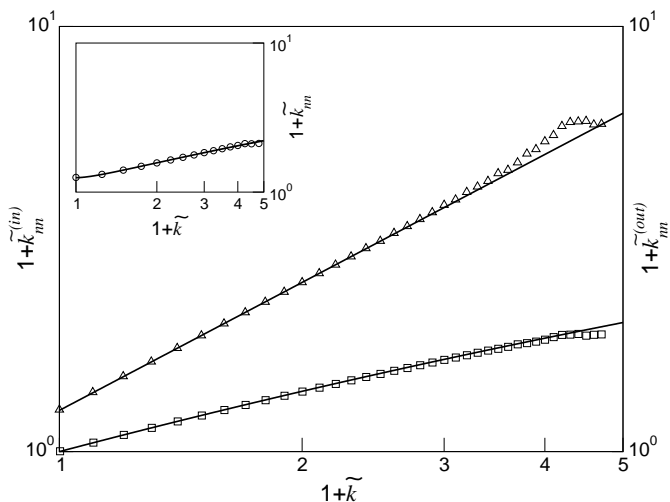} }
\caption{Histogram of $\wt{k}_{nn}^{(in)}$ (squares),
$\wt{K}_{nn}^{(out)}$ (triangles), as functions of $\wt{k}$ for
$t=10^4$, from simulations of the LPA model with $m=100$ and
$\alpha=500$ ($\beta <1/2$), and averaged over $10^4$
realizations. The behavior of $\wt{k}_{nn}$ is shown in the inset. The
solid lines are the analytic calculations.}
\label{BA_Knnk2}
\end{figure}

\begin{figure}
\centerline{\psfig{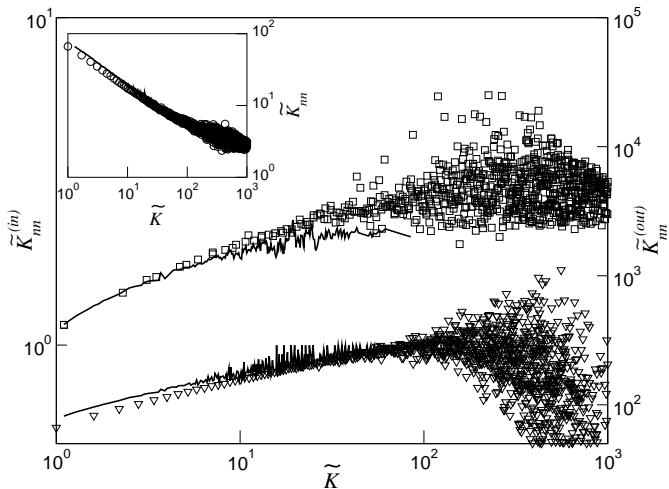}} 
\caption{Histogram of $\wt{K}_{nn}^{(in)}=\wt{k}_{nn}^{(in)}+1$
(squares), $\wt{K}_{nn}^{(out)}=\wt{k}_{nn}^{(out)}+1$ (triangles),
as functions of $\wt{K}=\wt{k}+1$ for $t=10^4$, from simulations of
the BB model with $m=10$, and averaged over $10^4$ realizations. The
behavior of $\wt{K}_{nn}=\wt{k}_{nn}+1$ is shown in the inset. The
solid lines are histogram from integration of the analytic expression.}
\label{BB-undir-Knn}
\end{figure}

Let now turn our attention to the BB model \cite{bianconi01}. The BB
model was proposed as a realistic model for the WWW, and represents a
paradigm for disassortatively mixed networks \cite{pastor01}. Here,
the preferential attachment mechanism is modified to embody the
intrinsic heterogeneity of nodes. This is done by assigning to each
node $j$ a quenched random variable, or {\it fitness}, $\eta_j$. The
network is grown by adding a node at each time step and connecting it
to $m$ existing nodes chosen with probability proportional to both
their degree and fitness $p_j(i+1) \propto \eta_j (k_j(i)+m)$.  Now
$k_j(i)$ depends on the single history of the network, and on the
quenched variables $\left\{ \eta_l \right\}_{l=1}^{i}$. However, for
any given realization of the quenched disorder the degree can be
approximated by $k_j(t)\simeq m\left( (t/j)^{\eta_j/c}-1\right)$,
where $c$ is a constant that depends on the probability distribution
of the fitness \cite{bianconi01}. Thus, even though $k_j(t)$ is a
function of all fitness, it essentially depends only on the value of
the fitness at site $j$. This approximation is found to be very
accurate numerically, and we will use it in what follows. Also, we
approximate $p_j(t)$ by replacing the normalization factor
$\sum_{l=1}^i \eta_l (k_l(i)+m)$ with its average value $mci$
\cite{barabasi99}.  In the same notations as above, we will measure
quantities in units of $m$.  The average degree of downstream
neighbors is given by $\wt{k}_{nn}^{out}(i)(t) = \sum_{j=1}^{i-1}
\Avg{\frac{\eta_j}{c i} \wt{k}_j(i) \wt{k}_j(t)}$; similarly the
average degree of upstream neighbors is $\wt{k}_{nn}^{in}(i)(t,\eta_i)
= \frac{\eta_i}{c \wt{k}_i(t) }\sum_{j=i+1}^{t} \Avg{\frac{\wt{k}_i(j)
\wt{k}_j(t)}{j}}$, where brackets represent the average over $\eta_j$.
Using the above approximations and computing the averages, one gets an
expression for these quantities as functions of $i$ and $\eta_i$. The
$\wt{k}$ dependence of $\tilde{k}_{nn}$ is then obtained by selecting
couples $(i,\eta_{i})$ that give rise to a degree $\wt{k}$ after $t$
steps, which can be sampled numerically. The results for a uniform
distribution of fitness in $[0,1]$ are shown in
Fig. \ref{BB-undir-Knn} where they are compared with results from
direct simulations.  Also in this case the degree of a node is
positively correlated with the average degree of both upstream and
downstream neighbors. However, as shown by Pastor--Satorras et
al. \cite{pastor01},  the nearest neighbors average degree decreases as a
function of the degree. 

In summary, we have demonstrated the crucial role of link directions
in the analysis of mixing patterns in complex networks, by showing
that assortativity patterns are often reversed once a network is
considered as directed.  In the growing complex network models we have
analyzed, we find positive correlations between the degree of a node
and the average degrees of both upstream and downstream nodes, while
fictitious correlations emerge when the different nature of the nodes
is not taken into account. This is an example of the Simpson's paradox
that may occur any time data from different sources are pooled
together. The correlation that appears in the pooled data is spurious:
a positive correlation between two quantities before pooling results
negative after pooling and vice versa.  In the particular case of
growing networks the degrees of upstream and downstream neighbors of a
node are positive correlated with the degree of the node itself,
however the correlation with upstream neighbors is much weaker. For
increasing degrees, the fraction of weakly correlated neighbors
increases. The overall neighbors' average degree can then decrease
as a result of the varied proportion, misleadingly suggesting the
presence of negative correlations.  In the case of BA networks, this
effect exactly balances that of positive correlations.\\
Our findings suggest the need for more detailed analysis of real
directed networks, such as the WWW, with a special focus on the
direction of links between nodes. The counterintuitive properties
described above may explain the anomalous exclusion of the WWW from
the realm of social networks based on its observed disassortative
mixing.

We thank Miguel--Angel Mu\~noz for useful discussions.


\begin{thebibliography}{99}
\bibitem{nota1} The average over realizations of the degree of a node
$i$ grows in time as $\wt{K}_i(t)=\wt{K}_i(i)f(t)/f(i)$, where
$f(t)=\Gamma(t+1/2)/\Gamma(t)$.  For large $t$ this gives the
asymptotic $\wt{K}_i(t)\simeq \wt{K}_i(i)c(i)\sqrt{t/i}$.  Each node
$i>1$ appears at time $t=i$ with a degree equal to $m$, so that
$\wt{K}_i(i)=1$. For $i=1$ it is convenient to choose $\wt{K}_1(1)=2$,
so that $\sum_{i=1}^t\wt{K}_i(t)=2t$. The coefficient $c(i)$ is
sensibly different $1$ (value in the continuum time approximation)
only for $i=1$, for which $c(1)=2/\sqrt{\pi}\simeq1.128\dots$, and we
will approximate it with one for all other values, so that finally
$\wt{K}_i(t)\simeq \sqrt{t/i}$ for $i>1$, while $\wt{K}_1(t)\simeq
4/\sqrt{\pi}\sqrt{t}$. With this initial conditions
$A=e^{8/\sqrt{\pi}}/\sqrt{2}$.

Analogous calculations can be performed for the LPA model, where  
$f(t)$ is replaced by $\Gamma(t+\beta)/\Gamma(t)$. This gives 
$A=(2^{1-2\beta}+(2\beta-1)/(1-\beta)^2)^{\beta/(2\beta-1)}$.
\end{thebibliography}
\end{document}